\documentclass[aps,prl,twocolumn,superscriptaddress]{revtex4-2}
\usepackage{graphicx}
\usepackage{color}

\newcommand{\overbar}[1]{\mkern 1.5mu\overline{\mkern-1.5mu#1\mkern-1.5mu}\mkern 1.5mu}

\usepackage{mhchem}

\begin{document}

\title{Charge transfer-induced Lifshitz transition and magnetic symmetry breaking in ultrathin CrSBr crystals}

\author{Marco Bianchi}
\affiliation{Department of Physics and Astronomy, Interdisciplinary Nanoscience Center (iNANO), Aarhus University, 8000 Aarhus C, Denmark}
\author{Kimberly Hsieh}
\affiliation{Department of Physics and Astronomy, Interdisciplinary Nanoscience Center (iNANO), Aarhus University, 8000 Aarhus C, Denmark}
\author{Esben Juel Porat}
\affiliation{Department of Physics and Astronomy, Interdisciplinary Nanoscience Center (iNANO), Aarhus University, 8000 Aarhus C, Denmark}
\author{Florian Dirnberger}
\affiliation{Institute of Applied Physics and W{\"u}rzburg-Dresden Cluster of Excellence ct.qmat, Technische Universit{\"a}t Dresden, Germany}
\author{Julian Klein}
\affiliation{Department of Materials Science and Engineering, Massachusetts Institute of Technology, Cambridge, Massachusetts 02139, USA}
\author{Kseniia Mosina}
\author{Zdenek Sofer}
\affiliation{Department of Inorganic Chemistry, University of Chemistry and Technology Prague, Technickaá 5, 166 28 Prague 6, Czech Republic}
\author{Alexander N. Rudenko}
\affiliation{Institute for Molecules and Materials, Radboud University, 6525 AJ Nijmegen, the Netherlands}
\author{Mikhail I. Katsnelson}
\affiliation{Institute for Molecules and Materials, Radboud University, 6525 AJ Nijmegen, the Netherlands}
\author{Yong P. Chen}
\affiliation{Department of Physics and Astronomy, Interdisciplinary Nanoscience Center (iNANO), Aarhus University, 8000 Aarhus C, Denmark}
\affiliation{Department of Physics and Astronomy and Purdue Quantum Science and Engineering Institute, Purdue University, West Lafayette IN 47907 USA}
\author{Malte R\"osner}
\email{m.roesner@science.ru.nl}
\affiliation{Institute for Molecules and Materials, Radboud University, 6525 AJ Nijmegen, the Netherlands}

\author{Philip Hofmann}
\email{philip@phys.au.dk}
\affiliation{Department of Physics and Astronomy, Interdisciplinary Nanoscience Center (iNANO), Aarhus University, 8000 Aarhus C, Denmark}
\date{\today}

\begin{abstract}
Ultrathin CrSBr flakes are exfoliated  \emph{in situ} on Au(111) and Ag(111) and their electronic structure is studied by angle-resolved photoemission spectroscopy. The thin flakes' electronic properties are drastically different from those of the bulk material and also substrate-dependent. For both substrates, a strong charge transfer to the flakes is observed, partly populating the conduction band and giving rise to a highly anisotropic Fermi contour with an Ohmic contact to the substrate. The fundamental CrSBr band gap is strongly renormalized compared to the bulk. The charge transfer to the CrSBr flake is substantially larger for Ag(111) than for Au(111), but a rigid energy shift of the chemical potential is insufficient to describe the observed band structure modifications. In particular, the Fermi contour shows a Lifshitz transition, the fundamental band gap undergoes a transition from direct on Au(111) to indirect on Ag(111) and a doping-induced symmetry breaking between the intra-layer Cr magnetic moments further modifies the band structure.
Electronic structure calculations can account for non-rigid Lifshitz-type band structure changes in thin CrSBr as a function of doping and strain. In contrast to undoped bulk band structure calculations that require self-consistent $GW$ theory, the doped thin film properties are well-approximated by density functional theory if local Coulomb interactions are taken into account on the mean-field level and the charge transfer is considered. 
\end{abstract}
\maketitle

The physics of two-dimensional (2D) magnetism is fascinating in its own right \cite{Mermin:1966uj,Chakravarty:1989aa,Jongh:1990aa,Irkhin:1999aa,Gibertini:2019aa}, and the recent realisation of 2D magnetic materials not only gives ample opportunity to investigate the fundamental principles of 2D magnetism, it also provides the technology for integrating magnetism into 2D materials-based devices via proximity exchange effects \cite{Zollner:2019aa}. Prominent examples of exfoliated 2D ferromagnets are CrI$_3$ \cite{Huang:2017aa}, Cr$_2$Ge$_2$Te$_6$ \cite{Gong:2017aa}, Fe$_3$GeTe$_2$ \cite{Deng:2018aa} and CrSBr \cite{Goser:1990aa,Telford:2020tc}. 
CrSBr stands out because of a high Curie and  N\'{e}el temperatures for intra-layer ferromagnetic and inter-layer antiferromagnetic ordering, respectively (150 and 132~K)\cite{Goser:1990aa,Telford:2020tc}. Its magnetic properties are stabilized by an orthorhombic crystal structure that has been theoretically predicted to lead to highly anisotropic bands in the lowest conduction bands, with strongly one-dimensional (1D)-like characteristics \cite{Guo:2018wt,Jiang:2018ug,Wang:2019wf,Wilson:2021ty,Klein:2023aa,Bianchi:2023aa}. The theoretical description of the bulk properties is highly challenging due to the need to account for magnetism (including disorder and paramagnetism in the high temperature phase), electronic correlations and long-range Coulomb interactions \cite{Bianchi:2023aa,rudenko_dielectric_2023}.

So far, the predicted anisotropic electronic states in the conduction band (CB) have only been observed indirectly via optical and transport properties in the bulk and down to single layers \cite{Telford:2020tc,Klein:2022aa,Telford:2022uy}. The bulk band structure in the paramagnetic phase has been determined by angle-resolved photoemission spectroscopy (ARPES) \cite{Bianchi:2023aa} but this does not give access to the CB in a semiconductor. 

Here we apply the recently introduced  kinetic  single-layer synthesis  (KISS) technique \cite{Cabo:2023} to prepare ultrathin flakes of CrSBr on Au(111) and Ag(111) substrates \emph{in-situ} with an atomically clean interface. Surprisingly, this leads to a giant charge transfer to the CrSBr flakes and a partial CB filling. This gives rise to an Ohmic contact between the materials, makes the anisotropic lowest CB directly accessible to ARPES measurements, and it induces an intra-layer magnetic symmetry breaking. Given their metallicity, the thin films can also be measured at low temperatures, solving the problem of charging that can be an issue for magnetic semiconductors \cite{Strasdas:2022,Bianchi:2023aa}. 

CrSBr crystals are synthesized by chemical vapor phase growth \cite{Klein:2022aa}.  ARPES experiments are performed at the SGM-3 beamline of ASTRID2 \cite{Hoffmann:2004aa}. The sample temperature is  35~K. The synchrotron radiation polarisation and the sample-to-analyzer direction are both in the plane of incidence and the analyzer slit is perpendicular to the plane of incidence.  Au(111) and Ag(111) films on mica are cleaned by noble gas sputtering and annealing. Surface cleanliness is judged by low energy electron diffraction and the quality of the electronic surface state in ARPES. Ultrathin CrSBr films are prepared using the KISS method  \cite{Cabo:2023}. In short, CrSBr crystals are cleaved in vacuum to obtain a clean surface and this is then gently pushed against  the cleaned metal substrate surface, leaving ultrathin flakes on the substrate. The resulting flakes are studied by optical microscopy, atomic force microscopy and ARPES. For additional experimental details, 
see Appendix.

To theoretically describe the doped flakes, we apply density functional theory (DFT) calculations for bilayer samples taking the local Coulomb interaction into account on the mean-field level within DFT+U~\cite{liechtenstein_density-functional_1995}, using $U=3.68$~eV and $J=0.39$~eV as recently estimated from constrained random phase approximation calculations~\cite{rudenko_dielectric_2023}. The charge transfer is modelled by artificially  adding  $+0.125$ electrons per layer, which is counter-acted by a positively charged background. All calculations are spin-resolved and are performed within the Vienna Ab Initio Simulation Package~\cite{kresse1,kresse2} applying the generalized gradient approximation within PBE~\cite{pbe} and using 16$\times$12 Monkhorst-Pack $k$-grids and an energy cut off 500~eV, starting from a layered anti-ferromagnetic ordering. The lattice constants and atomic positions are relaxed until forces acting on individual atoms are smaller than 5~meV/\AA\ whereby we take van der Waals effects into account~\cite{klimes_chemical_2009,klimes_van_2011} (for further details see Appendix).

\begin{figure}
\includegraphics[width=0.5\textwidth]{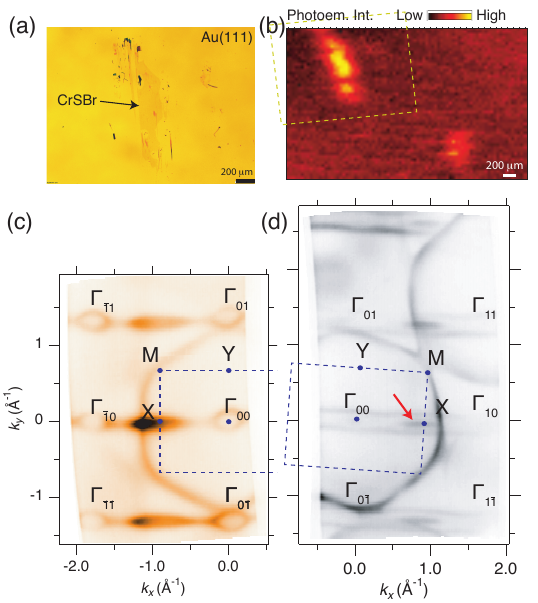}\\
  \caption{(Color online) (a) Exfoliated ultrathin CrSBr flake on Au(111) seen by optical microscopy. (b) The same flake imaged by ARPES as the photemission intensity of Br 3d core levels measured at 130~eV photon energy  (c) Photoemission intensity at the Fermi level for the same flake measured at 100~eV photon energy, showing a superposition of CrSBr and Au(111) features. The dashed lines mark the rectangular two-dimensional Brillouin zones of CrSBr. (d) Corresponding photoemission intensity at the Fermi level for a CrSBr flake deposited on Ag(111). The arrow indicates the band splitting at X, corresponding to Figure \ref{fig:dispersion_th}(g).}
  \label{fig:optical_fs}
\end{figure}

Figure~\ref{fig:optical_fs}(a) shows an optical microscopy image of a flake exfoliated on Au(111). The flake covers a large area of $\approx$800$\times$200$~\mu$m$^2$. It is almost transparent, with dark holes in the substrate clearly visible through the flake (such holes are a typical defect for Au(111) on mica). The thickness of the flake varies between 1 and 10~nm whereas flakes exfoliated on Ag(111) have a minimal thickness of 10~nm. The exact thickness does not seem to be critical because the observed electronic structure of the flakes does not vary across many preparations.  In addition to the very thin flakes studied here, the microscopic images typically also show large and much thicker flakes with the characteristic rectangular shape of CrSBr \cite{Klein:2022aa}. These appear black in the images and show the same charging effects in low-temperature ARPES that are also observed for bulk crystals \cite{Bianchi:2023aa}  
(for details see Appendix).

 A photoemission image of the same flake as in Fig.~\ref{fig:optical_fs}(a) is shown in Fig.~\ref{fig:optical_fs}(b), obtained by scanning the synchrotron radiation light spot across the sample while monitoring the photoemission intensity of Br 3d core levels. The flake is clearly visible but its boundaries appear blurred. This is due to the rather large light spot footprint ($\approx$60$\times$120~$\mu$m$^2$) that also leads to the CrSBr photoemission intensity always being observed together with that of the substrate. 

The photoemisison intensity at the Fermi level $E_\mathrm{F}$ for ultrathin CrSBr flakes on Au(111) and Ag(111) is shown in Figure~\ref{fig:optical_fs}(c) and (d), respectively. The CrSBr signal is easily distinguished from that of the substrates by symmetry alone. CrSBr has a rectangular unit cell and highly directional features corresponding to the expected periodicity are observed for both surfaces. The substrates, by contrast, give rise to the large hexagonal structures that represent cuts through the bulk Fermi surface. Observing a CrSBr Fermi contour implies that the flakes become metallic upon placing them on the substrates, creating an Ohmic contact. Note that the CrSBr features are aligned differently to the Au(111) and Ag(111) bulk Fermi surfaces. This alignment can be changed deliberately by choosing the desired relative orientation of the crystals in the KISS process. We do not observe any major changes of the CrSBr electronic structure as a function of relative CrSBr-substrate orientation, suggesting that the CrSBr-substrate interaction is dominated by screening and charge transfer rather than by hybridization or moir\'e potentials. Examples for differently oriented flakes are given in the 
Appendix.

\begin{figure}
 \includegraphics[width=0.5\textwidth]{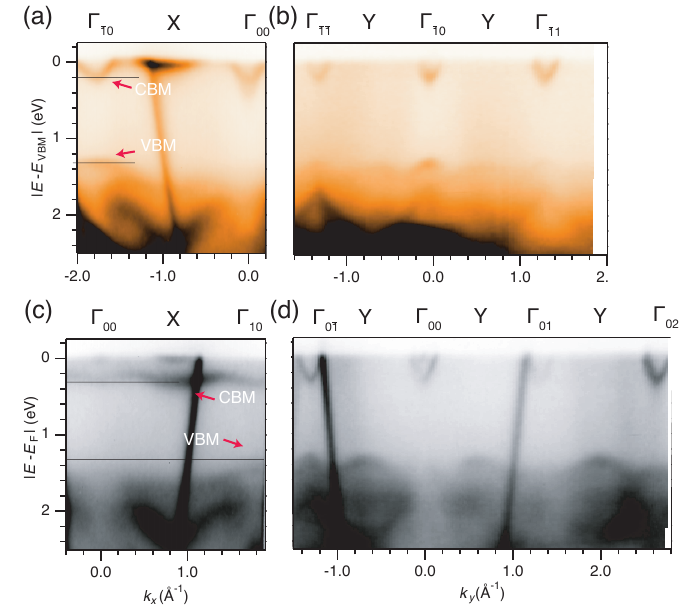}\\
  \caption{(Color online) Photoemission intensity along high-symmetry lines for ultrathin CrSBr on Au(111) and Ag(111) collected with $h\nu$=56~eV. (a), (b) Situation for Au(111) in the $x$ and $y$ direction, respectively. The conduction band minimum (CBM) and valence band maximum (VBM) are marked by arrows. The corresponding energies are indicated by horizontal lines. (c), (d) Corresponding data for Ag(111).}
  \label{fig:dispersion_exp}
\end{figure}

While the CrSBr flakes show quasi-1D Fermi contours on both substrates, the Fermi contour topology is surprisingly different. The origin of this can be understood by inspecting the dispersion of the CrSBr states for both substrates in the extended zone scheme in Fig.~\ref{fig:dispersion_exp}. Comparing the results along $\Gamma$-X in Fig.~\ref{fig:dispersion_exp}(a) and (c), it becomes clear that the electronic structure is similar on both substrates but the flakes are more strongly electron-doped on Ag(111). An electron-like band (with a positive effective mass) is observed around the $\Gamma$ point of the Brillouin zone (BZ) for CrSBr on both Au(111) and Ag(111). It corresponds to the bulk conduction band minimum (CBM) of CrSBr which is populated by charge transfer from the substrate. This band is more strongly populated on Ag(111). The binding energy maximum at $\Gamma$ is at 308$\pm 8$ and 187$\pm 8$~meV on Ag(111) and Au(111), respectively. This trend is consistent with the lower work function of Ag(111) leading to a stronger charge transfer \cite{Hansson:1978ab,Chelvayohan:1982aa} and it has also been observed for other 2D semiconductors \cite{Dendzik:2017ab}.

The electronic structure difference between the two substrates goes beyond a doping-induced rigid shift of the bands. Indeed, it resembles a Lifshitz-type transition of the band structure -- known earlier as electronic topological transition \cite{Lifshitz:1960aa,Lifshitz:1973,Blanter:1994aa,Katsnelson:1994aa} -- and even changes the fundamental band gap of the material. The evolution of the Fermi contours in Figure \ref{fig:optical_fs}(c) and (d) already shows the hallmark of Lifshitz transitions, namely, the disconnection of some Fermi contours and the appearance of others. The complexity of the changes can be fully appreciated when comparing the band structure change at the X point to that at $\Gamma$. For CrSBr on Au(111), the conduction band at X is barely populated with a maximum binding energy of 39$\pm 8$~meV. On Ag(111) the state is found at 317$\pm 8$~meV, corresponding to a much bigger shift than for the state at $\Gamma$. Indeed, the shift is so big that the conduction band at X ends up (just) below that at $\Gamma$, turning CrSBr on Ag(111) into an indirect band gap semiconductor. Given the fact that a strong hybridization of CrSBr and substrate states appears unlikely due to symmetry and that we do not find experimental evidence for this either, it is tempting to ascribe this band structure change to the different doping levels. We shall confirm this in the theoretical treatment below. 

Already a superficial inspection of the Fermi contours in Figure~\ref{fig:optical_fs}(c) and (d) appears to confirm the quasi-1D character of the CrSBr conduction band. For CrSBr / Au(111) the entire Fermi contour is comprised of the lowest conduction band. The contour is almost 1D, apart from the oval shape around $\Gamma$. On Ag(111), a second band above the CBM is partly populated, resulting in the photoemission maxima at the X points in Fig.~\ref{fig:optical_fs}(d). The lowest conduction band no longer crosses $E_\mathrm{F}$ in the X direction. This is the reason for the open parallel lines in Fig.~\ref{fig:optical_fs}(d). The dispersion around the conduction band minimum at $\Gamma$ is highly anisotropic for both surfaces. This can be seen from the effective masses obtained by fitting the detailed dispersion. These are 0.73$m_e$ (0.28$m_e$) and 1.3$m_e$ (0.2$m_e$) for the $x(y)$ direction on Au(111) and Ag(111), respectively 
(see Appendix).
The rather big difference of the effective masses between the two substrates is not surprising but merely a manifestation of the non-rigid band structure changes upon doping. 

The valence band maximum (VBM) is observed at the  $\Gamma$ points for CrSBr flakes on both Au(111) and Ag(111) in Figs.~\ref{fig:dispersion_exp}(a) and (b) (marked by arrows). The VBM is not equally well observed for every $\Gamma$ point in the extended zone scheme. In particular, the band is completely absent for normal emission at $\Gamma_{00}$. This has also been observed for the VBM in bulk samples where it has been ascribed to sub-lattice interference effects \cite{Bianchi:2023aa}. The band gap at $\Gamma$ determined for CrSBr flakes on Au(111) and Ag(111) is 1.14$\pm 0.03$~eV  and 1.18$\pm 0.02$~eV, respectively, much smaller than the bulk band gap of $\approx$2.1~eV \cite{Bianchi:2023aa}. This trend is similar to that observed for other 2D semiconductors on metallic substrates \cite{Dendzik:2017ab,Eickholt:2018aa,Antonija-Grubisic-Cabo:2015aa}. 

\begin{figure*}
  \includegraphics[width=0.99\textwidth]{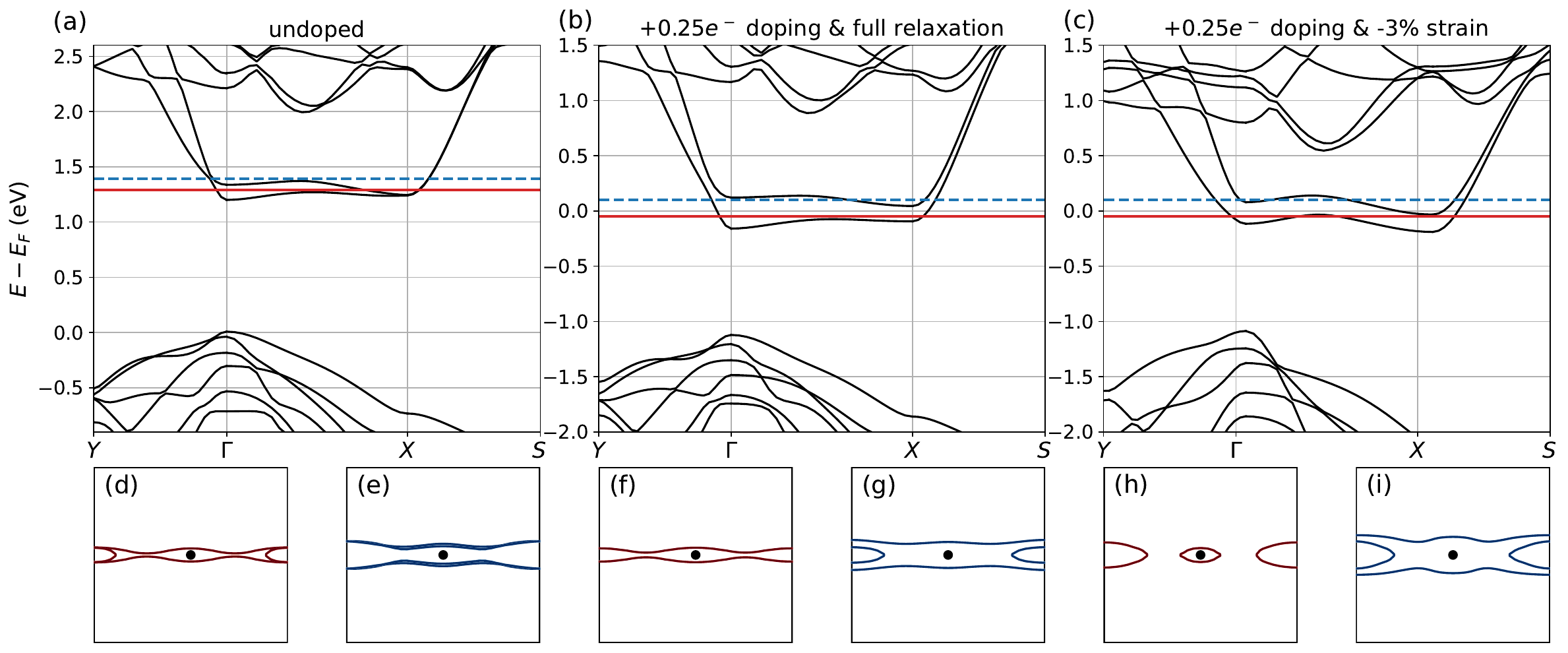}
  \caption{CrSBr Bilayer DFT+U band structures and Fermi contours. (a) Band structures without doping; (b) with a doping of $+0.125$ electron per layer; (c) with doping and applying in-plane compressive strain. (d) to (i) Effective Fermi contours resulting from these band structures according to the chemical potentials indicated by red and blue-dashed lines. In the doped cases the indicated chemical potentials are shifted by $-0.05$ and $+0.1$ eV w.r.t to the DFT+U predicted $E_\mathrm{F}$ to further illustrate the effects of modifications to the doping.}
  \label{fig:dispersion_th}
\end{figure*}

All of the experimental observations including their variations with substrates can be explained by comparison to our calculations.  As a minimal  model, we show bilayer band structures without doping, with electron doping, as well as with electron doping and under in-plane compressive strain in Fig.~\ref{fig:dispersion_th}(a) to (c). The strained case has been added for further comparison, keeping in mind that the lattice constants from our DFT+U calculations under electron doping might not be accurate. In the panels below the band structures we depict the accompanying Fermi contours to compare to the ARPES results. In all of these anti-ferromagnetic band structures, we find two strongly anisotropic Cr $d_{x^2-y^2/z^2}$-orbital-dominated lowest CBs, well separated from the other CBs and well defined band gaps on the order of $1$ to $1.3$~eV. Especially the latter points towards the validity of the simplified DFT+U approach here, which we understand as a result of the enhanced substrate and internal screening.

Without taking charge transfer effects into account, we find a direct band gap of about $1.3\,$eV at the $\Gamma$ point and the two CBs are degenerate at X. In such a scenario any finite doping (approximated by a rigid shift of the Fermi level into the CBs, as indicated by red and blue dashed lines) reaching the X minimum would create two Fermi contours with a full pocket around X for intermediate doping, c.f. Fig.~\ref{fig:dispersion_th}(d). 

Upon taking the charge transfer explicitly into account, the Cr atoms in each layer host different magnetic moments of $\pm2.87\, \mu_\mathrm{B}$ and $\pm2.78\,\mu_\mathrm{B}$, respectively, which is accompanied by a small charge disproportionation between these Cr atoms. As a result of this intra-layer magnetic symmetry breaking, the degeneracy at X is lifted, as shown in Fig.~\ref{fig:dispersion_th}(b). The resulting splitting is directly observable in the Fermi contour, as soon as the second band at X gets occupied, see transition between Fig.~\ref{fig:dispersion_th}(f) and (g).

Finally, upon applying an in-plane strain of $-3\%$, we see that the bands between $\Gamma$ and X become more dispersive and the CB minimum shifts to X yielding an indirect band gap. In this case, a small chemical potential can result in two fully detached pockets around $\Gamma$ and X, as indicated in Fig.~\ref{fig:dispersion_th}(h). Increasing the chemical potential gives again rise to a tube-like Fermi contour together with a detached second pocket around X, as shown in Fig.~\ref{fig:dispersion_th}(i).

Our theoretical results and especially their variations upon changing the chemical potential are in good qualitative agreement with the experimental findings. For the Au(111) supported flakes, a very low doping level is present with the CBM at $\Gamma$ being clearly occupied and the remaining CB straddling $E_\mathrm{F}$. This is similar to the situation in Figs.~\ref{fig:dispersion_th}(d) and (f) but there is no possibility to establish if the two bands at X are degenerate or not. For CrSBr flakes on Ag(111), on the other hand, the higher doping shows clear evidence of a band splitting at X, as seen, e.g., by the similarity of the calculated Fermi contours in  Figs.~\ref{fig:dispersion_th}(g) and (i) and the experimental one in Fig.~\ref{fig:optical_fs}(d) where the splitting is indicated by an arrow. This splitting underlines that the doping of ultra thin CrSBr does not just lead to a rigid displacement of the Fermi level but leads to a Lifshitz transition accompanied by the breaking of the intra-layer magnetic symmetry. Finally, the experimentally observed relocation of the absolute CBM from $\Gamma$ to X could be achieved by a small strain of the CrSBr film or be a small doping-dependent change of the equilibrium lattice constant. Note that a rather high compressive strain has been chosen here to exaggerate the relative shifts between the CB minima at $\Gamma$ and X.

The most interesting qualitative difference between our theoretical band structures and the experimental ARPES data is the rather deep electron pocket around $\Gamma$. Our DFT+U calculations cannot fully reproduce these characteristics, which might hint towards further band structure renormalization beyond the mean-field ones discussed here, such as electron-plasmon or electron-magnon coupling or further lattice structure renormalization effects. 

In conclusion, upon investigating ultra thin CrSBr flakes on Au(111) and Ag(111) substrates with clean interfaces obtained by \emph{in situ} exfoliation, we could systematically study the effects of charge transfer from the metallic substrates to the CrSBr layers. We found that this charge transfer induces non-rigid shifts and Lifshitz-like transitions in the CrSBr conduction bands. At small doping levels, we find a quasi 1D Fermi contour with an anisotropic electron pocket around  $\Gamma$, while at large doping the Lifshitz transition yields a full and non-closed tube-like pocket expanding through the entire Brillouin zone, accompanied by a second fully detached pocket around X. The comparison to mean-field DFT+U calculations including the effects of charge transfer allows us to interpret these characteristics as the result of an unexpected intra-layer magnetic symmetry breaking.  

\begin{acknowledgments}
We thank  Bjarke Rolighed Jeppesen for technical support. 
This work was supported by VILLUM FONDEN via the Centre of Excellence for Dirac Materials (Grant No. 11744), the Villum Investigator Program (Grant. No. 25931) and the Independent Research Fund Denmark  (Grant No. 1026-00089B). 
MR and MIK acknowledge the research program “Materials for the Quantum Age” (QuMat) for financial support. This program (registration number 024.005.006) is part of the Gravitation program financed by the Dutch Ministry of Education, Culture and Science (OCW). 
MIK is further supported by the ERC Synergy Grant “FASTCORR” (project 854843): Ultrafast dynamics of correlated electrons in solids.
FD gratefully acknowledges financial support from Alexey Chernikov and the W\"urzburg-Dresden Cluster of Excellence on Complexity and Topology in Quantum Matter ct.qmat (EXC 2147, Project-ID 390858490). JK gratefully acknowledges financial support from Frances Ross and the Alexander von Humboldt foundation.
Z.S. was supported by ERC-CZ program (project LL2101) from Ministry of Education Youth and Sports (MEYS) and used large infrastructure (reg. No. CZ$.02.1.01/0.0/0.0/15_0003/0000444$ financed by the EFRR).
\end{acknowledgments}

\section{Appendix}          

\subsection{Sample preparation and characterization}

Au(111) and Ag(111) thin film on mica ($\approx$200 nm thickness, Phasis, and $\approx$300~nm thickness, Georg-Albert-PVD, respectively) are cleaned by repeated cycles of Ne$^{+}$ sputtering (4$\times 10^{-6}$~mbar, 0.5~keV, 15~min) and annealing (15~min at 800~K). Bulk CrSBr crystals are cleaved in ultra high vacuum using scotch tape and characterized by ARPES, giving same results as reported in Ref.  \cite{Bianchi:2023aa}. Flakes are deposited at room temperature and at 180-190~K on Au(111) and Ag(111), respectively, using the KISS method described in Ref. \cite{Cabo:2023}. Different orientations can be obtained by rotating the bulk CrSBr crystals against the substrate before applying the KISS method. 

\begin{figure}
 \includegraphics[width=0.5\textwidth]{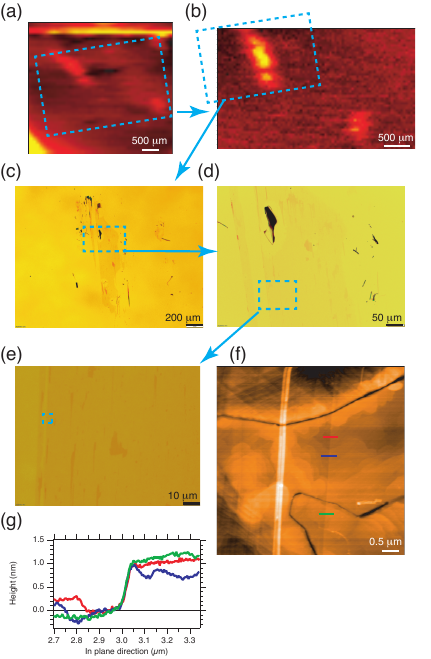}\\
  \caption{Topographical characterization of the measured CrSBr flake deposited on Au(111). (a) Large-scale map showing the photoemission intensity at $E_\mathrm{F}$ (b) Small-scale map of a flake, showing the photoemission intensity around the Br 3d core levels. (c) - (e) Corresponding optical images of the flake with increasing magnification. (f) AFM measurement on the side of the flake marked by a blue rectangle in panel (e). (b) Height profiles across the edge of the deposited flake and along the coloured lines in panel (f).}
  \label{fig:MicroscopeAu}
\end{figure}

\begin{figure}
 \includegraphics[width=0.5\textwidth]{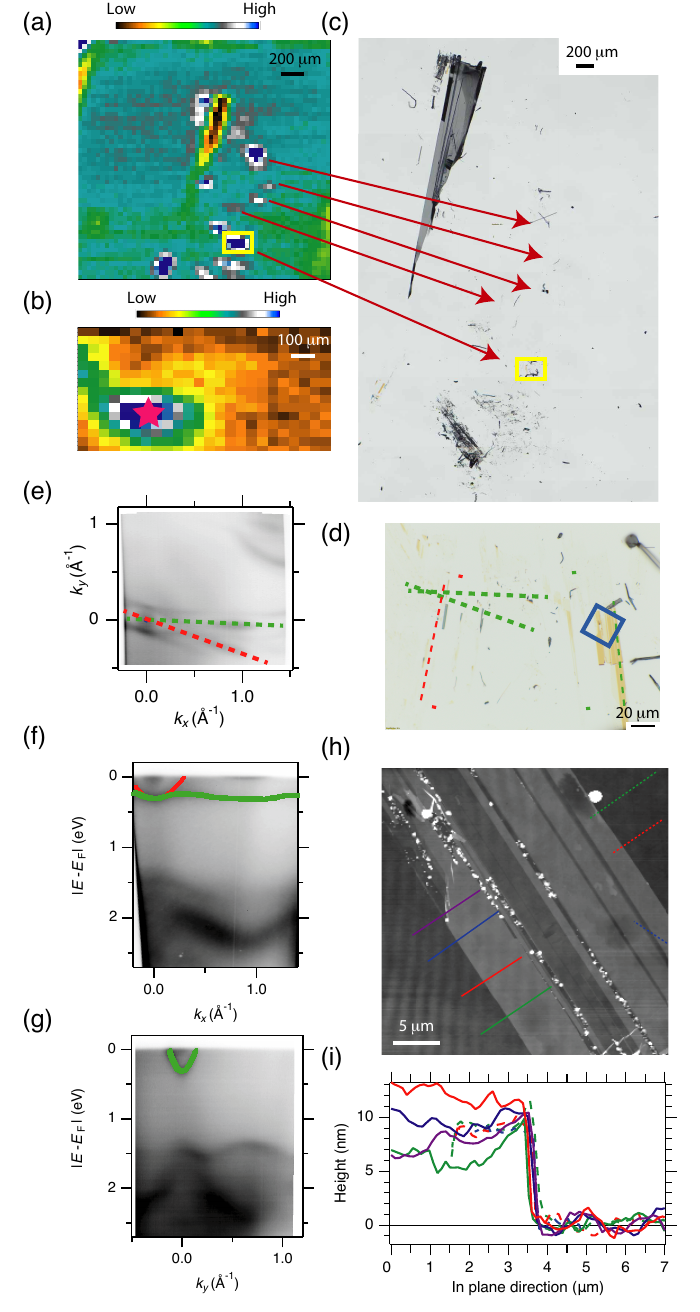}\\
  \caption{Topographical characterization of the measured CrSBr flake deposited on silver and measured in ARPES. (a) Large and (b) small scale photoemission maps of the flakes, respectively. (b) Area around the yellow rectangle in (a). The red star indicates where the band structure measurements for the main text have been performed. (c) Optical image of the sample surface. The yellow rectangle marks the  investigated area in ARPES. (d) Magnification of the area within the yellow rectangle in panel (c). (e) Photoemission intensity at $E_\mathrm{F}$ at 56~eV  photon energy; (f), (g) correspondent dispersion along $k_x$ and $k_y$, respectively. Red and green lines in (e) mark the orientation of the two rotational domains that can be recognized in the high resolution microscopic image in (d). Corresponding colours in (f) and (g) indicate the photoemission contributions from the two different domain orientations. (h) AFM topography in the area marked by the blue rectangle in panel (d). (i) Height profiles along the coloured lines in panel (h). }
  \label{fig:MicroscopeAg}
\end{figure}

Figures \ref{fig:MicroscopeAu} and \ref{fig:MicroscopeAg} show microscopic characterizations of the measured CrSBr flakes on Au(111) and Ag(111), respectively, using photoemission intensity maps, optical microscopy and atomic force microscopy (AFM). For a characterisation by photoemission, it turns out to be advantageous to map the large-scale structure using the photoemission intensity at $E_\mathrm{F}$, as this reliably shows the holes in the substrate and is insensitive to non-crystalline Br contaminations arising from the KISS process.  Fig.~\ref{fig:MicroscopeAu}(a) shows such a large-scale photoemission intensity map at $E_{\mathrm{F}}$. Once promising structures are identified, it is advantageous to map the photoemission intensity around the Br 3d core levels to achieve maximum contrast between the flake and the Au(111) background on a smaller scale. This is illustrated in  Fig.~\ref{fig:MicroscopeAu}(b). Images of the same flake acquired with an optical microscope with increasing magnifications are shown in panels (c) to (e). As pointed out in the main text, the CrSBr flake is almost transparent with the large (black) holes in the substrate clearly seen through the flake. The area marked by a blue rectangle in panel (e) was used for an AFM characterisation given in panel (f). Here the edge of the flake is visible as a faint vertical line through the image. Height profiles perpendicular to this flake edge along the coloured lines are shown in panel (g) and indicate a step height of about 1~nm. Different measurements acquired in other areas of the flake reach a maximum step height of about 10~nm.

Figure \ref{fig:MicroscopeAg} shows the corresponding results for the Ag(111) substrate.  Panels (a) and (b) shown the large-scale and zoomed-in photoemission intensity maps at $E_\mathrm{F}$ and at the Br 3d core level energy, respectively, with panel (b) being a zoom-in area around the flake used for the ARPES measurements reported in the main text (the exact position for these measurements is given by the star). Panel (c) shows an overview by optical microscopy with arrows between panels (a) and (c) indicating the corresponding islands. The yellow rectangle marks the area in which ARPES data were collected. Interestingly, the large elongated CrSBr flake in the upper left part of the images shows a very low photoemission intensity in panel (a), in contrast to all the other flakes. This is due to this being a very thick flake suffering from charging. Fig.~\ref{fig:MicroscopeAg}(d) shows a magnification of the area used for the ARPES measurements and reveals the presence of several flakes with slightly different orientations. This, combined with the large photon beam, makes it very challenging to single out one flake for band structure measurements. The measurements in Fig.~1(d) of the main text are evidently dominated by a single flake because there is only one orientation of CrSBr Fermi contour visible. For the band structure measurements reported in Fig.~2(c) and (d) of the main text, however, there is a small contamination from a different CrSBr flake present. The origin of this is illustrated in Fig.~\ref{fig:MicroscopeAg}(e) which shows the photoemission intensity at $E_\mathrm{F}$ taken with a photon energy of 56~eV, as used for mapping the dispersion in Fig.~2(c) and (d) of the main text. In this case, two Fermi contours are visible, slightly rotated against each other (their orientations are given by the red and green dashed lines). Likely candidates for creating these Fermi contours are seen near the  dashed lines of the same colour in Fig.~\ref{fig:MicroscopeAg}(d). Corresponding dispersions along $k_x$ and $k_y$ are shown in panels (f) and (g), with the signal from the two different domains marked in green and red. The absence of the red contamination signal in Fig.~1(d) of the main text can be explained by the higher photon energy that leads to a slightly smaller and shifted photon beam.

AFM measurements performed in the area marked by the blues square in Fig. \ref{fig:MicroscopeAg}(d) are shown in panel (h) where the lines indicate the path for the height profiles given in panel (i). In the case of CrSBr flakes on Ag(111), the minimum step height observed on the measured flake is 10~nm.  The roughness of the substrates makes the precise determination of the thickness difficult, however, and given the limited spatial resolution of the photon beam, it is reasonable to assume that the photoemission intensity stems from areas with varying thickness of a few nm for both substrates, all appearing transparent in optical microscopy.

\subsection{ARPES measurements}

The energy resolution of the ARPES measurements was better than 25 and 60~meV for  measurements with 56 and 100~eV photon energy, respectively. The angular resolution was better than 0.2$^{\circ}$. We have measured the photon energy dependence of the observed CrSBr band positions in order to explore a possible three-dimensional character of the band structure in the flakes of finite thickness. No such photon energy dependence was observed. 

\begin{figure}
 \includegraphics[width=0.5\textwidth]{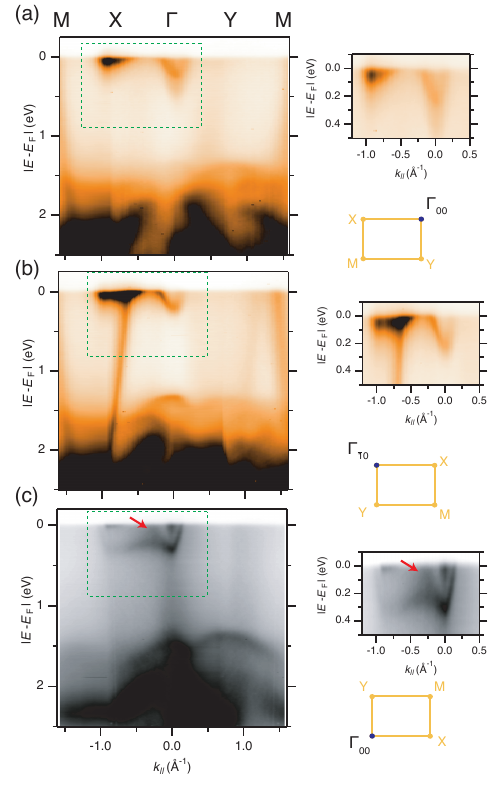}\\
  \caption{Photoemission intensity along high-symmetry lines for ultrathin CrSBr on Au(111) and Ag(111) measured at 56~eV photon energy. (a), (b) Situation for Au(111) for two equivalent but different paths in the extended zone scheme (indicated on the right hand side).(c) Corresponding data for Ag(111). Red Arrows in (c) indicate contribution to the photoemission intensity coming from a second flake with different azimuthal orientation shown in Fig.~\ref{fig:MicroscopeAg}. }
  \label{fig:dispersion2_exp}
\end{figure}

As as supplement to Fig.~2 of the main text, a more detailed view on the CrSBr band structures on Au(111) and Ag(111) is given in Fig.~\ref{fig:dispersion2_exp}. Panels (a) and (b) show the detailed dispersion of CrSBr CB on Au(111) along two equivalent but different paths in the extended zone scheme. Note that the dispersion around normal emission $\Gamma_{00}$ appears smeared out with a substantial intensity at higher binding energy that is absent around $\Gamma_{\bar{1}0}$ in panel (b). This is due to the contribution of the Au(111) surface state. Normal emission data is thus unsuitable to obtain high-quality band structure data from the CrSBr conduction band on Au(111). The same issue exists on Ag(111) as seen by the peak near the Fermi energy in Fig.~\ref{fig:dispersion2_exp}(c). Here, however, the shallower binding energy of the surface state and the higher binding energy of the CrSBr band lead to a large separation between the states.

\begin{figure}
 \includegraphics[width=0.5\textwidth]{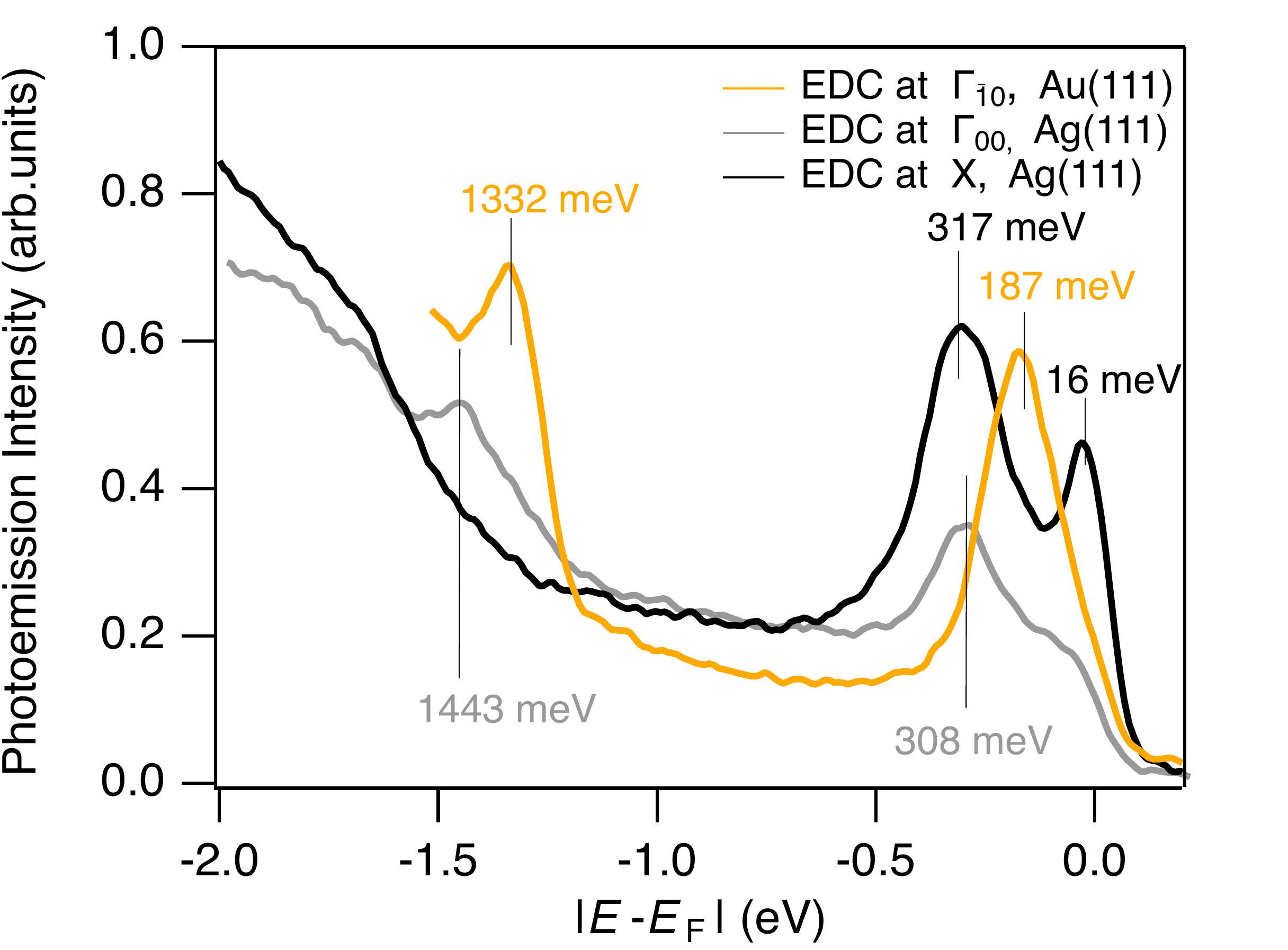}\\
  \caption{Energy distribution curves at selected high symmetry points of ultra thin CrSBr on Au(111) (yellow) and Ag(111) (black and gray), respectively. The energy positions of the peaks given in the figure are obtained by a fit.}
  \label{fig:EDC}
\end{figure}

The energy positions of the band extrema for CrSBr are determined by inspecting energy distribution curves (EDCs) at high symmetry points. Such an analysis is shown in Fig.~\ref{fig:EDC}. Values on peak positions are obtained by fitting each single EDC with a Lorentzian line shape and a polynomial background, multiplied by a Fermi-Dirac distribution. The uncertainty in the peak position is $\approx$8~meV. In order to obtain more accurate values for the energy gaps reported in the main text, EDCs around the high symmetry points are analyzed and the energy position of the band maximum (or minimum) is determined by a fit of the resulting energy positions by a second order polynomial.

\begin{figure}
 \includegraphics[width=0.4\textwidth]{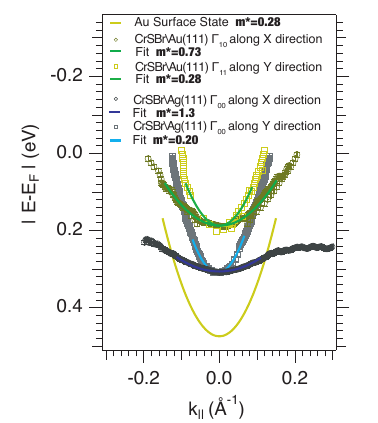}\\
  \caption{Comparison between the CB dispersion at $\Gamma$ for the two high symmetry directions along $k_x$ and $k_y$ for CrSBr flakes on Au(111) and Ag(111). As a reference the Au(111) surface state dispersion is plotted with the parameters taken from Ref. \cite{Nechaev:2009aa}  but without Rashba splitting.}
  \label{fig:EffectiveMasses}
\end{figure}

A similar EDC-based analysis combined with a quadratic interpolation of the peak positions near the high symmetry points can be used to determine the effective mass  $m^{\ast}$  at the band extrema. This is illustrated in Fig.~\ref{fig:EffectiveMasses} for different substrates and directions. As the considered bands are not necessarily parabolic, this method leads to an uncertainty on $m^{\ast}$ that depends on the energy range that is considered for the fit. This energy range is chosen to be  100~meV measured from the band minimum. The uncertainties are up to 40\% of the obtained value. For comparison, the Au(111) surface state dispersion is also shown in Fig.~\ref{fig:EffectiveMasses}, as obtained from the Ref. \cite{Nechaev:2009aa} but without Rashba-splitting.

\begin{figure}
 \includegraphics[width=0.5\textwidth]{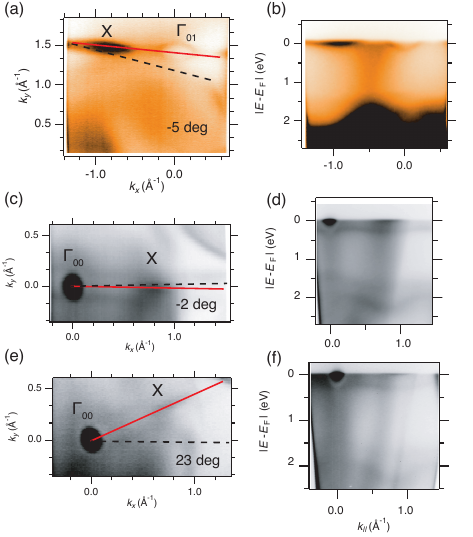}\\
  \caption{ARPES measurements on differently oriented ultrathin CrSBr flakes on Au(111) and Ag(111). (a) Photomission intensity at $E_\mathrm{F}$ for a CrSBr flake on Au(111) with $\Gamma-X$ direction (red line) rotated by 5$^\circ$ away from the $\overbar{\Gamma}\overbar{K}$ direction of the substrate (dashed line). (b) Dispersion along the red line. (c)-(f) corresponding data for CrSBr flakes on Ag(111) rotated by -2$^\circ$ and 23$^\circ$, respectively. }
  \label{fig:DifferentOrientation}
\end{figure}

No appreciable difference in band dispersion was observed for different orientations of the CrSBr flakes with respect to the substrate. This is illustrated in Fig.~\ref{fig:DifferentOrientation} with one orientation for a flake on Au(111) different from the main text and two additional orientations for Ag(111). The observed dispersion along $\Gamma-X$ shown in the right hand panels is the same as reported in the main text. 

\subsection{DFT lattice structure and magnetic anisotropy}

\begin{figure}
 \includegraphics[width=0.5\textwidth]{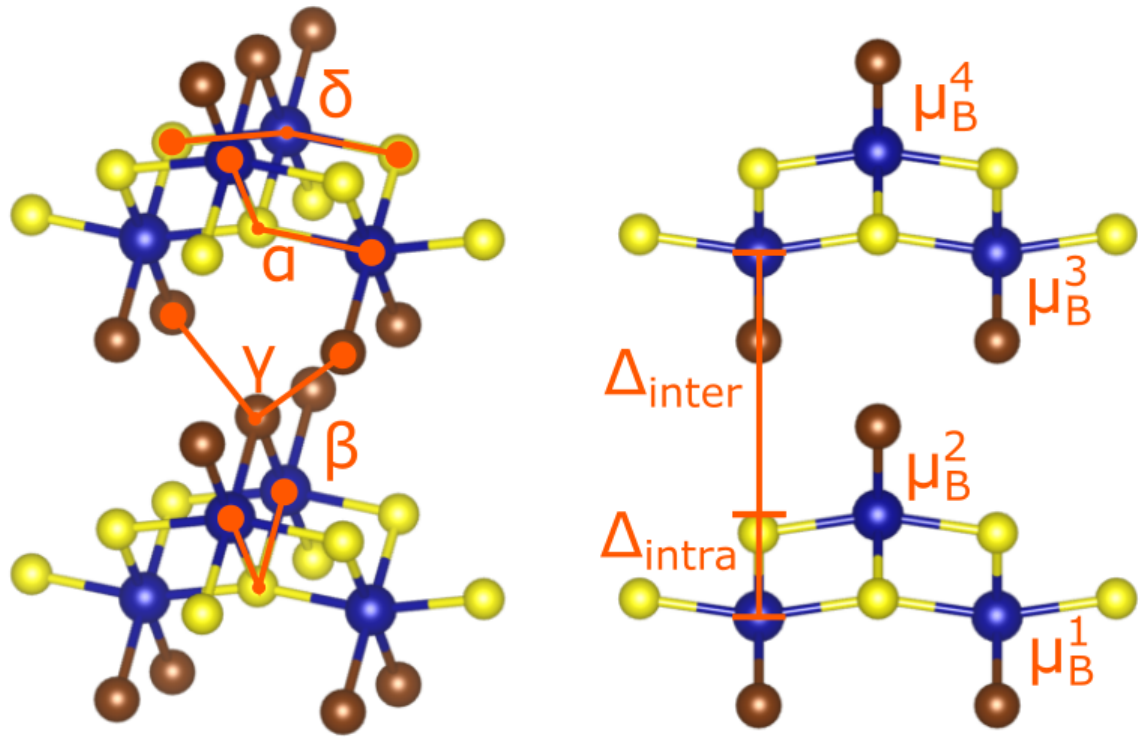}
  \caption{CrSBr bilayer lattice structure indicating analysed angles, distances, and magnetic moments. For values see Tab.~\ref{tab:DFTStructure}.}
  \label{fig:DFTStructure}
\end{figure}

In Fig.~\ref{fig:DFTStructure} and Tab.~\ref{tab:DFTStructure} we depict the structural and magnetic differences between the pristine (undoped) and doped CrSBr bilayer, as resulting from our DFT+U calculations. While all indicated lattice details slightly change, the main difference is the reduction of $\Delta_\text{intra}$, which is the Cr-Cr $z$-distance in the same layer, while $\Delta_\text{inter}$, the Cr-Cr $z$-distance between the nearest-neighbour Cr atoms of different layers, does not change. The individual layer thus get effectively thinner under doping. As a result we find the largest angle change in $\delta$. As the inter-layer distance stays the same, the inner and outer Cr atoms of the bilayer system experience a different change in their local environments, which we interpret as the reason for the intra-layer magnetic symmetry breaking as reflected in $\mu_B^1 \neq \mu_B^2$ and $\mu_B^3 \neq \mu_B^4$.

\begin{table}
    \centering
    \caption{CrSBr lattice structure details for undoped and doped bilayers including  angles, distances, and magnetic moments as indicated in Fig.~\ref{fig:DFTStructure}. \label{tab:DFTStructure}} 
    \begin{ruledtabular}
        \begin{tabular}{rrrc}
          & undoped & doped & difference \\
         \hline
         lat. const. $a$ [\AA] & 3.483 & 3.496 & -0.013 \\
         lat. const. $b$ [\AA] & 4.734 & 4.783 & -0.049 \\
         \hline
         $\alpha$ [$^\circ$] &95.84 &94.77 &1.07 \\
         $\beta$ [$^\circ$] &92.85 &93.34 &-0.50 \\
         $\gamma$ [$^\circ$] &120.01 &120.54 &-0.53 \\
         $\delta$ [$^\circ$] &163.31 &166.63 &-3.32 \\
         \hline
         $\Delta_\text{intra}$ [\AA] &2.00 &1.93 &0.08 \\
         $\Delta_\text{inter}$ [\AA] &5.21 &5.21 &0.00 \\
           \hline
         $\mu_B^1$ &2.77 &2.87 &-0.11 \\
         $\mu_B^2$ &2.77 &2.78 &-0.01 \\
         $\mu_B^3$ &-2.77 &-2.78 &0.01 \\
         $\mu_B^4$ &-2.77 &-2.87 &0.11 \\
        \end{tabular}
    \end{ruledtabular}
\end{table}

%

\end{document}